\documentclass[prx, aip, twocolumn, superscriptaddress, amsmath, amssymb, nofootinbib]{revtex4} 			 

\usepackage[utf8x]{inputenc}
\usepackage{graphicx}
\usepackage{bm} 
\usepackage{dcolumn} 
\usepackage[colorlinks=true, citecolor=blue, linkcolor=blue]{hyperref}%
\usepackage[normalem]{ulem}  

\newcommand{\cob}[1]{{\color{black}#1}}
\allowdisplaybreaks

\draft 

\begin{document}

\title{Nonlinear behaviour of the nonradiative exciton reservoir in quantum wells}

\author{A.~S.~Kurdyubov}
\affiliation{Spin Optics Laboratory, St. Petersburg State University, Ulyanovskaya 1, Peterhof, St. Petersburg, 198504, Russia}

\author{A.~V.~Trifonov}
\affiliation{Spin Optics Laboratory, St. Petersburg State University, Ulyanovskaya 1, Peterhof, St. Petersburg, 198504, Russia}
\affiliation{Experimentelle Physik 2, Technische Universit{\"a}t Dortmund, 44221 Dortmund, Germany}

\author{A.~V.~Mikhailov}
\affiliation{Spin Optics Laboratory, St. Petersburg State University, Ulyanovskaya 1, Peterhof, St. Petersburg, 198504, Russia}

\author{Yu.~P.~Efimov}
\author{S.~A.~Eliseev}
\author{V.~A.~Lovtcius}
\affiliation{Resource Center ``Nanophotonics'', St. Petersburg State University, Ulyanovskaya 1, Peterhof, St. Petersburg, 198504, Russia}

\author{I.~V.~Ignatiev}
\affiliation{Spin Optics Laboratory, St. Petersburg State University, Ulyanovskaya 1, Peterhof, St. Petersburg, 198504, Russia}
\date{\today}

\begin{abstract}
\cob{Excitons and free charge carriers with large wave vectors form a nonradiative reservoir, which can strongly affect properties of bright excitons due to the exciton-exciton and exciton-carrier interactions. In the present work, the dynamics of quasiparticles in the reservoir at different areal densities is experimentally studied in a GaAs/AlGaAs quantum well using a time-resolved reflectance spectroscopy of nonradiative broadening of exciton resonances. The population of the reservoir is controlled either by the excitation power or by the temperature of the structure under study. The dynamics is quantitatively analyzed in the framework of the model developed earlier [Kurdyubov {\em et al.,} Phys. Rev. B {\bf 104}, 035414 (2021)]. The model considers several dynamic processes, such as scattering of photoexcited excitons into the reservoir, dissociation of excitons into free charge carriers and the reverse process, carrier-induced exciton scattering into the light cone depleting the reservoir, thermally activated nonradiative losses of charge carriers. We have found that competition of these processes leads to highly nonlinear dynamics of reservoir excitons, although their density is far below the exciton Mott transition densities. Characteristic rates of the processes and their dependencies on the excitation power and the temperature are determined.}
\end{abstract}

\maketitle 

\section{Introduction}

\cob{Optically active or bright excitons dominate the optical properties of semiconductor heterostructures, especially at low temperatures. Bright excitons studied in real experiments are almost always surrounded by a cloud of dark excitons. The dark excitons with a large wave vector, $K_X$, exceeding the wave vector of light, $K_c$, live for a long time, tens of nanoseconds, in high-quality heterostructures~\cite{Trifonov2015}. Therefore, they can accumulate to huge densities, orders of magnitude higher than the density of bright excitons. Dynamic processes in the dark exciton reservoir strongly affect the dynamics of bright excitons due to the exciton-exciton interaction.}
This interaction plays particularly important role in formation of polaritons and their dynamics in quantum well (QW) structures with microcavities~\cite{Kavokin-book2017, Schmidt-PRL2019, Alyatkin-PRL2020, Vladimirova2010}. 

Experimental study of dark excitons is complicated because their direct detection in optical experiments \cob{in the visible and near infrared spectral regions} is impossible. Several indirect methods have been used for their study. In early works~\cite{Weisbuch-SSC1981, Damen-PRB1990, Deveaud-PRL1991, Szczytko-PRL2004, Deveaud-ChemPhys2005}, the formation mechanism of the bright excitons has been studied mainly via detection of the PL kinetics. It was found that the PL is observed within a few nanoseconds after the nonresonant pulsed excitation whereas the exciton radiative lifetime is only of about 10 picoseconds~\cite{Deveaud-PRL1991}. This observation indicates slow processes in the nonradiative reservoir, which, in particular, includes the coupling of the photocreated free electrons and holes into nonradiative excitons with large $K_X$, the energy relaxation (cooling) of the excitons, and their scattering into the light cone. The rate of these processes is strongly dependent on the exciton and carrier densities in the reservoir. In these experiments, a rapid PL rise within a few tens of picoseconds is also observed even at weak excitation~\cite{Damen-PRB1990, Szczytko-PRL2004, Robart-Amand-SSC1995}. 

These seemingly contradictory results have provoked extensive discussion about the nature of PL at the exciton optical transition~\cite{Szczytko-PRL2004, Deveaud-ChemPhys2005, Kira-PRL1998}. In particular, Kira {\em et al.}~\cite{Kira-PRL1998} have suggested a model of the Coulomb-correlated carriers, which are able to emit light with the exciton resonance photon energies just after their photocreation. This suggestion has stimulated development of new experimental methods of study of the nonradiative reservoir. 

In the general case, the nonradiative reservoir is a mixture of nonradiative excitons and the electron-hole plasma. The ratio of their densities depends on the sample temperature as well as on the excitation conditions~\cite{Szczytko-PRL2004, Deveaud-ChemPhys2005, Robart-Amand-SSC1995, Colocci-JAP1990, Scalbert1999}. To experimentally study the exciton and carrier dynamics in the reservoir the combined optical pump -- terahertz (THz) probe experiments have been performed in a number of works \cob{for QW heterostructures~\cite{Cerne-PRL1996, Kaindl-Nature2003, Chatterjee-PRL2004, Kappei-PRL2005, Kaindl-PRB2009} as well as for bulk crystals~\cite{Suzuki-PRB2011, Ulbricht-Review2011, Suzuki-PRL2012, Sekiguchi-PRB2015, Sekiguchi-PRL2017}}. The absorption of the THz radiation in the exciton subsystem occurs due to the resonant $1s \rightarrow 2p$ transition in the exciton. In the case of free carriers there is a continuous absorption spectrum. This difference in the spectra can be used, in principle, to determine the ratio of the exciton and carrier densities in the reservoir. Unfortunately, the relatively small cross-section of the THz absorption allowed one to perform such experiments only at relatively strong optical pumping \cob{when the exciton density in the reservoir becomes close to the exciton Mott transition density}. At such experimental conditions the dynamic equilibrium free carriers $\leftrightarrow$ excitons is shifted towards the carriers and their effective temperature increases. In other words, the experimental conditions for studying exciton dynamics become different from those typically used in most studies.

Recently we suggested a new method of experimental study of the dynamics of the excitons and free carriers in the reservoir~\cite{Trifonov-PRL2019, Kurdyubov-PRB2021}. Our method is based on the interaction of the reservoir quasi-particles with bright excitons. This method is experimentally \cob{implemented} by a pump-probe technique with spectral resolution. This technique allows one to obtain the dynamics of various parameters of exciton resonances observed in the reflectance spectra. For study of the exciton-exciton and exciton-carrier interaction the nonradiative broadening of the resonances is the most important parameter. We have already demonstrated experimentally in Ref.~\cite{Kurdyubov-PRB2021} and theoretically in Ref.~\cite{Gribakin-PRB2021} that the exciton resonance broadening provides direct information about the interaction of bright excitons with excitons and charge carriers in the reservoir. In contrast to the PL technique, the exciton polarization created by the probe pulses immediately detects the state of the reservoir created by the pump pulses. Together with the spectrally selective excitation, this technique provides rich information about the exciton and carrier dynamics in the reservoir.

Results of the study performed in Ref.~\cite{Kurdyubov-PRB2021} show that the dynamic processes in the reservoir are mainly controlled by the interaction of excitons and charge carriers. The exciton-exciton and exciton-phonon interaction play less important role. In particular, the PL dynamics is determined by the carrier-induced, rather than phonon-induced, scattering of excitons from the reservoir into the states within the light cone. The exciton and carrier dynamics depends both on the exciton and carrier densities. 

\cob{In the present work we study the dynamic processes via controlling the exciton and carrier densities. To change these densities we modify the excitation power and the sample temperature. We consider the case when the exciton densities are far below those for the exciton Mott transitions extensively studied in~Refs.~\cite{Kappei-PRL2005, Sekiguchi-PRL2017}. We have found that, despite the low exciton densities, the exciton and carrier dynamics in the reservoir is nonlinear and strongly depends on the excitation power}. The main source of the nonlinearity is the dependence on the carrier density squared and on the product of the exciton and carrier densities~\cite{Kurdyubov-PRB2021}. Besides, several effects contribute to this nonlinearity. They are the pump-induced decrease of the light absorption and the decrease of the exciton--exciton and exciton--free carrier scattering cross-sections. The pump-induced sample temperature rise shifts the dynamic equilibrium in the reservoir towards free charge carriers thus accelerating the dynamic processes. In addition, a channel of nonradiative losses of the charge carriers appears with the temperature rise in the structure under study. It results in a decrease of the PL yield at elevated temperatures. Analysis of dynamic processes in the framework of the basic model developed earlier~\cite{Kurdyubov-PRB2021} allowed us to quantitatively describe the experiments and to determine the power and temperature dependencies of the model parameters. 

\section{Resonant excitation} 
\label{Resonant-excitation}

\begin{figure}[ht]
\includegraphics[width=1\columnwidth]{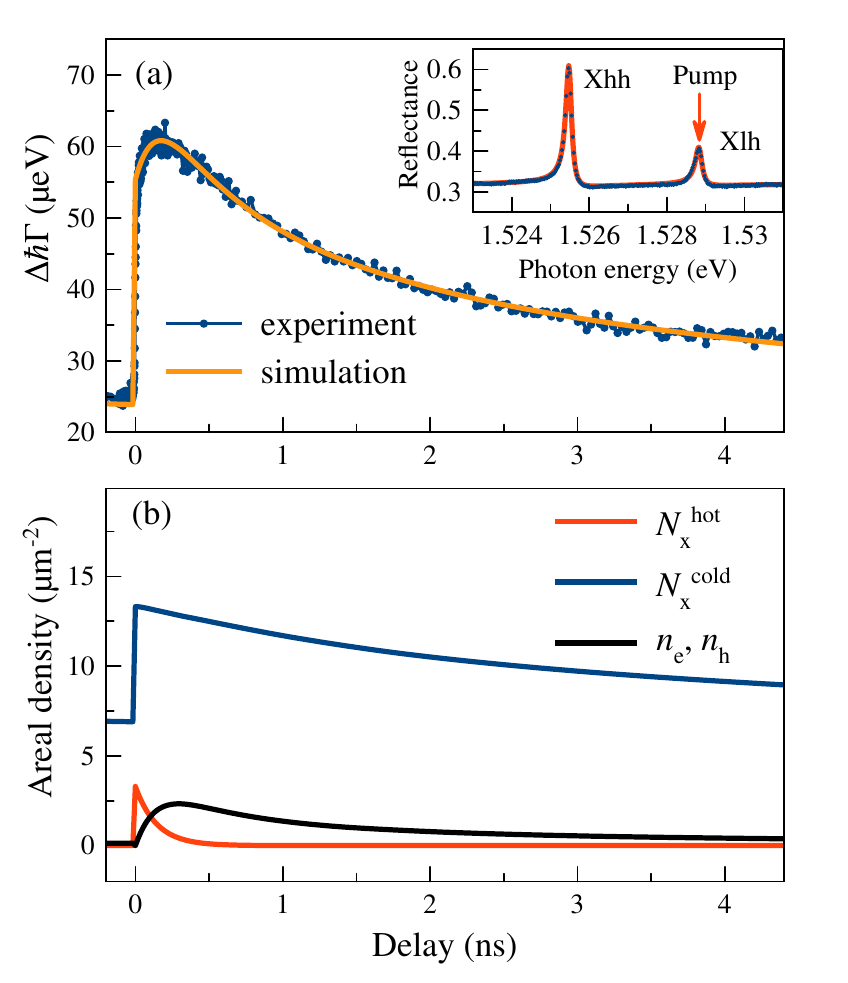}
    \caption{(a) Dynamics of the photoinduced nonradiative broadening $ \Delta \hbar \Gamma $ of the Xhh resonance under excitation into the Xlh resonance. Experimental data are shown by the blue dots, simulation by the red curve. The inset shows the reflectance spectrum with no pumping. (b) Dynamics of the areal densities of electrons, holes and excitons predicted by the model. Excitation power $ P_{\text{exc}} = 0.2$~mW. \cob{The laser spot diameter is 130~$\mu$m}. Sample temperature $T = 4.5$~K.
} \label{fig:dynamics_T670}
\end{figure}

The sample under study is described in detail in the previous work~\cite{Kurdyubov-PRB2021}. It contains one 14-nm GaAs/AlGaAs MBE-grown QW of high quality. The exciton reflectance spectrum of the sample is shown in the inset of Fig.~\ref{fig:dynamics_T670}(a). The heavy-hole (Xhh) and light-hole (Xlh) exciton resonances are observed in the spectrum as peaks due to appropriate choice of the top barrier layer thickness in the structure. The spectral widths of the resonances are determined by the radiative and nonradiative broadenings of the exciton transitions. The nonradiative broadening is mainly controlled by the exciton-exciton and exciton-carrier interactions. The inhomogeneous broadening of exciton resonances is found to be negligibly small, not exceeding a few tens of $\mu$eV, that confirms the high quality of the structure. 

The experimental technique is described in Ref.~\cite{Kurdyubov-PRB2021}. Here we briefly comment on it. A 80-femtosecond Ti-sapphire laser is used as a light source. The repetition rate of the laser pulses $\nu_r=80$~MHz. The laser beam is divided into the pump and probe beams. The pump beam is passed through a spectral selector, which reduces its spectral width at half maximum down to $\approx 1$~meV increasing the pulse duration up to $\approx 2$~ps. Such pump beam allows one to selectively excite the light-hole excitons. The spectrally broad probe beam is used to detect the reflection in the region of the Xhh and Xlh resonances at each delay between the pump and probe pulses. The spectra are accumulated by a spectrometer equipped by a CCD matrix detector. 

The measured spectra are analyzed using a model of non-local dielectric response of excitons in QWs~\cite{Ivchenko-book}. The amplitude reflection at the exciton resonance is described as:
\begin{equation}
r_{QW}(\omega) = \frac{i\Gamma_0}{(\omega_0 - \omega)-i(\Gamma_0 + \Gamma)}.
\label{Eqn:rQW}
\end{equation}
The intensity of the reflected light also depends on the amplitude reflection from the sample surface $r_s$,
\begin{equation}
R(\omega) = \left|\frac{r_s + r_{QW}(\omega)e^{i2\phi}}{1 + r_s  r_{QW}(\omega)e^{i2\phi}}\right|^2.
\label{Eqn:R}
\end{equation}
 Due to small inhomogeneous broadening, the model perfectly describes the exciton resonances that allows one to reliably determine their main parameters such as the exciton transition energy ($\hbar \omega_0$) and the radiative ($\hbar \Gamma_0$) and nonradiative  ($\hbar \Gamma$) broadenings. An example of the fit of the exciton resonances is shown in the inset of Fig.~\ref{fig:dynamics_T670}(a) by the red curve.

Fits of the resonances observed at different time delays between the pump and probe pulses provide time dependencies of all the parameters. Fig.~\ref{fig:dynamics_T670}(a) shows an example of the dependence of the photoinduced nonradiative broadening of the Xhh resonance under excitation into the Xlh resonance at a relatively low power of 0.2~mW (the pump spot diameter is 130~$\mu$m). The part of the broadening not related to the excitation is subtracted from these data. As seen, the broadening increases during of about 0.2~ns and then slowly relaxes within the time window experimentally studied. The rate of the relaxation continuously decreases with time so that the non-zero photoinduced broadening is observed even at ``negative'' delays. This is the broadening induced by the action of preceding pump pulses.

\begin{figure}[ht]
\includegraphics[width=1\columnwidth]{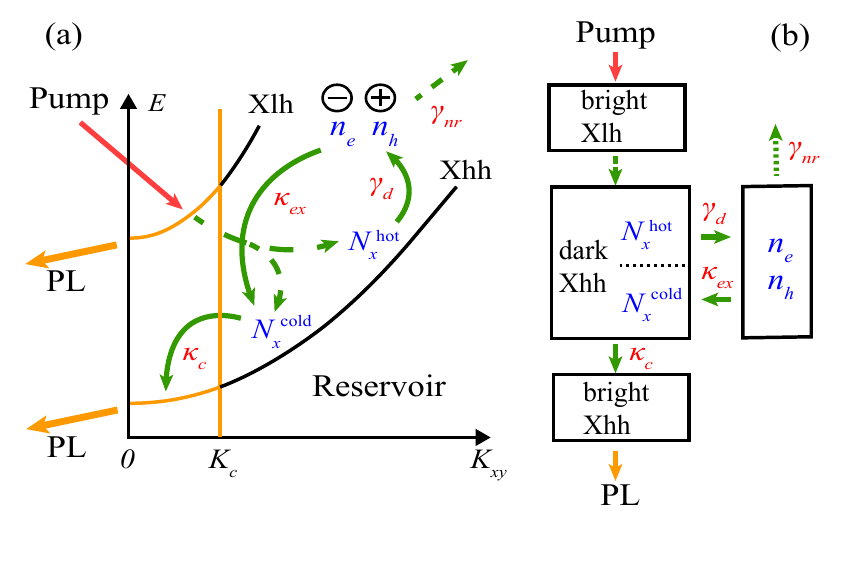}
    \caption{Scheme of the processes describing the dynamics of excitons and free charge carriers under excitation into the Xlh resonance. The processes are shown in the K-space (a) and in the flowchart (b).
} \label{fig:scheme1}
\end{figure}

The basic model of the broadening dynamics for the case of weak excitation is described in Ref.~\cite{Kurdyubov-PRB2021}. Although the model is very simplified, it takes into account the most important processes using a minimum number of free parameters and describes the experimentally measured dynamic curves quite well.  Here we consider more general model applicable for the case when the effective reservoir temperature can be increased by the sample heating or by the strong pumping. 
The model assumes that the resonant pumping excites exciton polarization that can either create the secondary emission (the reflected light) or can be scattered as excitons into the reservoir. 
\cob{The fraction $k$ of the scattered excitons acquires the average temperature of the reservoir. We call them ``cold'' excitons with the initial density $N_x^\text{cold} = k N_0$. Here $N_0$ is the full areal density of excitons scattered into the reservoir. Only ``cold'' excitons can be scattered into the light cone in our model. Other dark excitons with higher wave vectors and energies are ``hot'' excitons. They cannot be efficiently scattered into the light cone and can only dissociate into free electrons and holes with some rate $\gamma_d$. Correspondingly, their density, $N_x^\text{hot}$, exponentially decreases in time, $N_x^\text{hot} = (1 − k)N_0 \exp(−\gamma_d t)$.} The dissociation process explains the initial slow rise of the broadening seen in Fig.~\ref{fig:dynamics_T670}(a). 

The last two processes considered in the model are the coupling of free electrons and holes into excitons, characterized by the bimolecular coefficient $\kappa_{ex}$, and the scattering of excitons into the Xhh exciton state in the light cone by free charge carriers, characterized by the bimolecular coefficient $\kappa_{c}$. The Xhh excitons in the light cone recombine so fast ($\tau_{PL} \le 10$~ps) that we can neglect the back scattering of these excitons into the reservoir. In the case of the low sample temperature, we can neglect the nonradiative decay of excitons and free carriers due to the high quality of the structure under study~\cite{Kurdyubov-PRB2021}. We can also neglect the radiative recombination of free electrons and holes since the oscillator strength of this process is of about two order of magnitude less than that of exciton recombination~\cite{Szczytko-PRL2004, Deveaud-ChemPhys2005}. 

To reduce the number of free parameters, we do not consider the contributions of the exciton-exciton and exciton-phonon interactions to the scattering of excitons into the light cone. The presence of a long-lived reservoir indicates that these processes are less efficient than the exciton scattering by free charge carriers. The fact that exciton-exciton scattering into the light cone has low efficiency is intuitively clear. In order to the exciton momentum to decrease to a small value corresponding to the light cone, the exciton must experience a head-on collision with another exciton with the opposite momentum. This process is statistically unlikely. Scattering of the reservoir excitons by electrons into the light cone is more efficient, since electrons have a smaller mass, interact with each other due to Coulomb forces, and can collectively take away the momentum of the exciton.

The system of kinetic equations describing the considered processes reads:
\begin{equation}
\left\{
   \begin{aligned}
      \frac{dn}{dt} & = \gamma_d N_x^{\text{hot}} - \kappa_{ex} n^2,     \\
      \frac{dN_x^\text{cold}}{dt~~~} & = \kappa_{ex} n^2 - \kappa_{c} 2 n N_x^\text{cold}.
   \end{aligned}
\right.
   \label{Eqn-system1}
\end{equation}
Here $ n \equiv n_e = n_h $ are the densities of electrons and holes, which are assumed to be equal (there are no residual carriers of one sign); $N_x^\text{cold}$ and $N_x^\text{hot}$ are the densities of ``cold'' and ``hot'' excitons, respectively (\cob{see the definitions of ``cold'' and ``hot'' above in this section}). The initial conditions for these variables are: $ n(0) = 0 $, $ N_x^\text{cold}(0) = N_{bgr} + N_0 k $, where $ N_{bgr} $ is the background density of excitons accumulated in the reservoir due to the action of preceding laser pulses. The processes considered in the model are schematically illustrated in Fig.~\ref{fig:scheme1}. Note that the temperature-activated losses of charge carriers described by $\gamma_{nr}$ will be included in the model in Sect.~\ref{temperature}.

The system of equations~(\ref{Eqn-system1}) is nonlinear relative to the exciton and free carrier densities. Besides, parameters of the model also depend on the excitation power and, hence, on the densities. All this determines the nonlinear behavior of the reservoir. System~(\ref{Eqn-system1}) has no analytical solution and we solved it numerically. For the approximation of the experimental curves we used $ N_0 $, $ k $, $ \gamma_d $, $ \kappa_{ex} $, and $ \kappa_c $ as fitting parameters. The photoinduced nonradiative broadening of the resonance Xhh measured in the experiments can be expressed in terms of densities of excitons and free carriers as
\begin{equation}
   \Delta \hbar \Gamma = \sigma_{eh} (n_e + n_h) + \sigma_x (N_{x}^\text{hot} + N_x^\text{cold}).
   \label{scattering}
\end{equation}
Here $ \sigma_{eh} $ and $ \sigma_{x} $ are the exciton-electron (and also exciton-hole, X-eh) and exciton-exciton (X-X) scattering cross-sections, respectively. The main mechanism of scattering of excitons and free carriers at small wave vectors is exchange interaction, see~\cite{Ciuti-PRB1998, Ramon2003, Ouerdane2008, Gribakin-PRB2021}.
The analysis carried out in these works shows that the exchange constants for X-eh and X-X interactions in narrow QWs have close values. For simplicity, we put $ \sigma_{eh} = \sigma_{x} \equiv \sigma $~\cite{Comment-Honold1988}. The calibration of areal density of the photocreated excitons performed in Ref.~\cite{Kurdyubov-PRB2021} gives rise to value of $\sigma = 3$~$\mu$m$^2 \cdot \mu$eV. 

The theoretical curve obtained in the framework of this model well describes the experimentally observed dynamics, see Fig.~\ref{fig:dynamics_T670}(a). The model also predicts the dynamics of the exciton and carrier densities, which are shown in Fig.~\ref{fig:dynamics_T670}(b). The calibration procedure to get the absolute values of the densities is described in Ref.~\cite{Kurdyubov-PRB2021}. 
The parameters obtained from the modeling are: fraction of ``cold'' excitons $ k \approx 0.7 $, the dissociation rate $ \gamma_d = 7$~ns$^{-1}$, the bimolecular coefficients $\kappa_{ex} = 0.6$~$\mu$m$^2/$ns and $\kappa_{c} = 0.08$~$\mu $m$^2 / $ns. The obtained value of $\kappa_{ex}$ is of the same order of magnitude as in Refs.~\cite{Piermarocchi1996, Piermarocchi1997, Deveaud2005}.

\section{Temperature dependencies of resonance broadening}
\label{temperature}

\subsection{Experiment}
Control of the exciton and carrier densities in the reservoir can be easily implemented by variation of sample temperature. In Ref.~\cite{Trifonov2015}, we already demonstrated that the heating of the sample accelerates excitonic relaxation. This effect was explained by the temperature-induced dissociation of nonradiative excitons. Here we consider this effect in more detail.

\begin{figure}
\includegraphics[width=1\columnwidth]{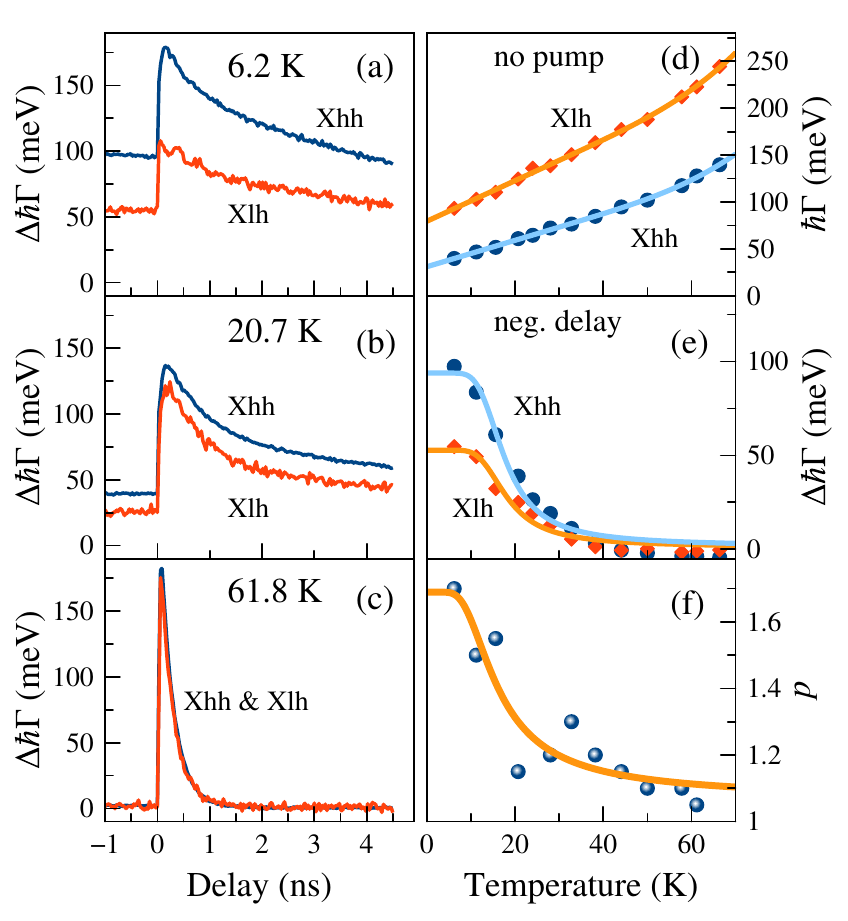}
\caption{(a - c) Dynamics of the photoinduced broadening $ \Delta \hbar \Gamma $ of the Xhh (blue lines) and Xlh (red lines) resonances under excitation into the Xhh \cob{with power $P = 1$~mW} at various sample temperatures. Background broadening not related to the optical pumping is subtracted. (d) Temperature dependencies of $ \hbar \Gamma $ of the Xhh and Xlh resonances (circles) in the absence of the pumping. Solid curves are the approximations by function~(\ref{Eq-phonons}). (e) Temperature dependencies of $ \Delta \hbar \Gamma $ for large (``negative'') delays. Solid curves are fits by the formula~(\ref{Eq-Arrhenius}). (f) Temperature dependence of ratio of nonradiatve broadenings of the Xhh and Xlh exciton resonances (circles). The solid curve is the fit by function~(\ref{pT}) with parameters: $p_0 = 0.75$, $E_{\text{act}} = 2.5$~meV, $a = 2.9$, $A_0 = 0.97$.
}
\label{T670_temp}
\end{figure}

The first thermal effect observed even in the absence of pumping is an increase of the nonradiative broadening $ \hbar \Gamma $ with the temperature increase. This dependence is shown in panel~(d) of Fig.~\ref{T670_temp} for the heavy-hole and light-hole excitons. It can be explained by the interaction of excitons with the acoustic and optical phonons (see~\cite{Lee1986, Poltavtsev2014}):
\begin{equation}
	\label{Eq-phonons}
\hbar\Gamma(T) = \hbar\Gamma(0) + \hbar \gamma_{ac} T + \frac{\hbar\Gamma_{\text{LO}}}{\exp[E_{\text{LO}}/(kT)]-1}.
\end{equation}

The energy of the optical LO-phonon was taken for GaAs: $ E_{\text{LO}} = 36 $~meV~\cite{Waugh-PR1963}. Fitting parameters for Xlh: $ \hbar \Gamma(0) = 81~\mu $eV, $ \hbar \gamma_{ac} = 2.2~\mu $eV/K, $ \hbar \Gamma_{\text{LO}} = 9.6$~meV; for Xhh: $ \hbar \Gamma(0) = 37~\mu$eV, $ \hbar \gamma_{ac} = 1.5~\mu$eV/K, $ \hbar \Gamma_{\text{LO}} = 6.5 $~meV. The obtained values of $ \hbar \gamma_{ac} $ and $ \hbar \Gamma_{\text{LO}} $ are close to those reported in Ref.~\cite{Poltavtsev2014}.

Panels (a - c) in Fig.~\ref{T670_temp} illustrate the dynamics of the $ \Delta \hbar \Gamma $ of the Xhh and Xlh resonances when the exciton system is pumped into the Xhh resonance at several temperatures. As before, for convenience, the pump-independent contribution shown in panel (d) is subtracted from the total broadening $ \hbar \Gamma $. Two main effects can be distinguished in the temperature dependencies. 

The first one is the shortening of the exciton lifetime in the reservoir when the temperature increases. That leads to a gradual decrease in the broadening $\Delta \hbar \Gamma$ at long (negative) delays. Panel (e) illustrates this effect. The experimental data can be well approximated by the Arrhenius curve:
\begin{equation}
	\label{Eq-Arrhenius}
Ar(T) = \frac{A_0}{1 + a e^{-E_{\text{act}}/ (k_B T)}}.
\end{equation}
Fitting the experimental data shown in Fig.~\ref{T670_temp}(e) by function (\ref{Eq-Arrhenius}) gives the following values of the free parameters for both curves (Xhh and Xlh): $ a \approx 65 $, $ E_{\text{act}} = 7$~meV. Parameter $E_{\text{act}}$ in the Arrhenius function has a physical meaning of a potential barrier for a temperature-activated process. The obtained value of $E_{\text{act}}$ coincides with the exciton binding energy (the exciton Rydberg) in the QW under study. The qualitative explanation of this effect is clear. At low temperature, the main process of the reservoir depopulation is the scattering of the excitons by free carriers into the light cone~\cite{Kurdyubov-PRB2021}. This process results in the decay of the nonradiative broadening of exciton resonance during the first several nanoseconds as it is shown in Fig.~\ref{fig:dynamics_T670}. During this time, the free electrons and holes couple into excitons and the scattering process terminates. Correspondingly, the remaining excitons survive in the reservoir for a long time (tens of nanoseconds) while less efficient processes, like exciton-exciton or exciton-phonon processes, slowly deplete the reservoir. However, when the structure heats up, the excitons become unstable and decay into free carriers, that accelerate the efficient exciton-carrier scattering process. This is why the activation energy of this process coincides with the exciton Rydberg energy.

Another effect observed in Fig.~\ref{T670_temp}(a-c) with increasing temperature is the decrease in the difference between dynamics of $ \Delta \hbar \Gamma $ for the Xhh and Xlh excitons. As it is already discussed in Ref.~\cite{Kurdyubov-PRB2021}, the photoinduced broadening of the Xlh resonance is approximately half of that of the Xhh resonance at low temperatures [see Fig.~\ref{T670_temp}(a)]. The reason for that is related to peculiarities of interaction of the reservoir excitons with bright ones. The main mechanism of such interaction is the exchange by electrons and holes~\cite{Ciuti-PRB1998, Gribakin-PRB2021}. At low temperatures, the reservoir contains mainly the Xhh excitons whose interaction with the bright Xhh excitons occurs via exchange by both electrons and holes. However the interaction with the bright Xlh excitons can occur only via exchange by electrons because the holes are of different types. 

At elevated temperatures the reservoir excitons can populate the Xlh states, which are 3.3~meV higher than the Xhh states [see inset in Fig.~\ref{fig:dynamics_T670}(a)]. The Xlh reservoir excitons can interact with the Xlh bright excitons via exchange by both the electrons and light holes. Thus, the interaction of the Xhh and Xlh bright excitons with reservoir excitons is leveled out. To quantitatively analyze this effect we have calculated the ratio of maximum values of nonradiative broadenings of the Xhh and Xlh exciton resonances shown in panels (a-c) of Fig.~\ref{T670_temp}, $p = [\Delta \hbar \Gamma(Xhh)]/[\Delta \hbar \Gamma(Xlh)]$. The temperature dependence of $p$ is shown in panel (f) of this figure. It can be approximated by function
\begin{equation}
p(T) = p_0 +Ar(T),
\label{pT}
\end{equation}
where $Ar(T)$ is the Arrhenius function~(\ref{Eq-Arrhenius}). The activation energy for this process is found to be $E_{\text{act}} = 2.5$~meV. This energy is close to the energy distance between the Xhh and Xlh states, that confirms the mechanism of leveling of the Xhh and Xlh resonance broadenings.

\subsection{Modeling}
\label{Modeling}

To model the experimental data shown in Fig.~\ref{T670_temp}, we should modify Eqs.~(\ref{Eqn-system1}). Namely, we should take into account that the elevated sample temperature activates processes of the reservoir depletion. We consider only one of such processes, the nonradiative loss of free carriers with the rate $\gamma_{nr} (T)$. 

The modified equation system reads:
\begin{equation}
\left\{
\begin{aligned}
&\frac{d n}{dt}  = \gamma_d N_{x}^\text{hot} - \kappa_{ex} n^2 - \gamma_{nr} (T) n, \\
&\frac{d N_x^\text{cold}}{dt~~~}  = \kappa_{ex} n^2 - \kappa_c 2n N_x^\text{cold}. \\
\end{aligned}
\right.
   \label{Eqn-system1T}
\end{equation}
Here $n$ is the areal density of the free electrons and holes and $N_x^\text{cold}$ is that of the ``cold'' excitons in the reservoir. The experimental data were obtained at relatively strong pumping, $P = 1$~mW, per a small spot with a diameter of about 100~$\mu$m. Therefore we assume that the laser pulses create the hot excitons only. In other words, the fraction of cold excitons $k=0$. A detailed discussion of the power dependence of parameter $k$ is presented in the next section. The hot excitons dissociate into free electrons and holes with characteristic rate $\gamma_d$: $ N_{x}^\text{hot} = N_0 e ^{-\gamma_{d} t}$, where $N_0$ is the exciton density in the reservoir created by a single laser pulse. The initial conditions are: $n(0) = 0$; $N_x^\text{cold}(0) = N_{bgr}$, where  $N_{bgr}$ is the density of cold excitons created by the preceding laser pulses.  It is determined from the value of $\Delta\hbar\Gamma_{\text{neg}}$ at negative delays,
$$
\Delta\hbar\Gamma_{\text{neg}} = \sigma N_{bgr},
$$
using the value of the cross-section of the exciton-exciton scattering $\sigma = 3$~$\mu$eV$  \cdot \mu$m$^2$~\cite{Kurdyubov-PRB2021}.

\begin{figure}[ht]
\includegraphics[width=1\columnwidth]{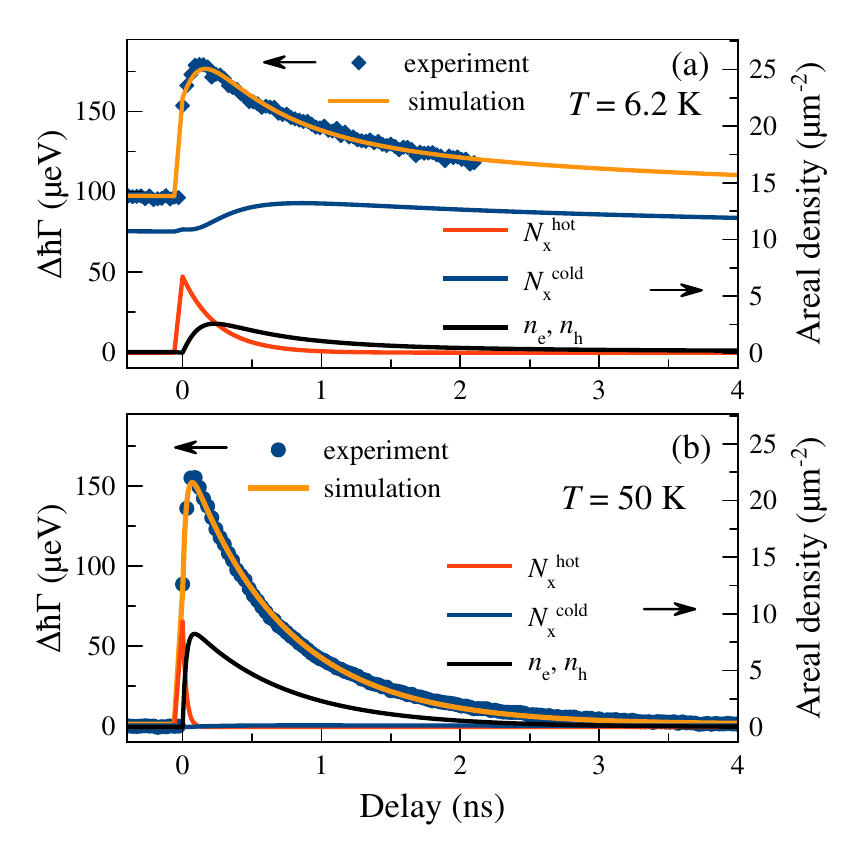}
    \caption{(a) Dynamics of the nonradiative broadening of the Xhh resonance under excitation into this resonance. The excitation power $P = 1$~mW. The laser spot diameter $d \approx 100$~$\mu$m. The sample temperature $T = 6.2$~K. Dots are the experimental data, the orange solid curve is the simulation (left axis). The color curves are the areal densities of the excitons and carriers in the reservoir (right axis). (b) The same for $T = 50$~K. } \label{fig:temp}
\end{figure}

The model has been applied to simulate the dynamics of the pump-induced nonradiative broadening of the Xhh resonance at different temperatures in the range 6~-- 60~K. Examples of the simulated dynamic curves are shown in Fig.~\ref{fig:temp}. As seen the simulation well reproduces the experimental curves. The model also allows us to simulate the dynamics of the areal densities of excitons and free carriers in the reservoir. The simulation shows that the density of the hot excitons decays within $\approx 1$~ns at $T = 6.2$~K and much faster at $T = 50$~K. The density of free carriers simultaneously increases with approximately the same rate and then decays at much longer time. The density of the cold excitons, compared to the densities of free carriers, is relatively large at low temperature and negligibly small at $T = 50$~K. 

\begin{figure}[ht]
\includegraphics[width=1\columnwidth]{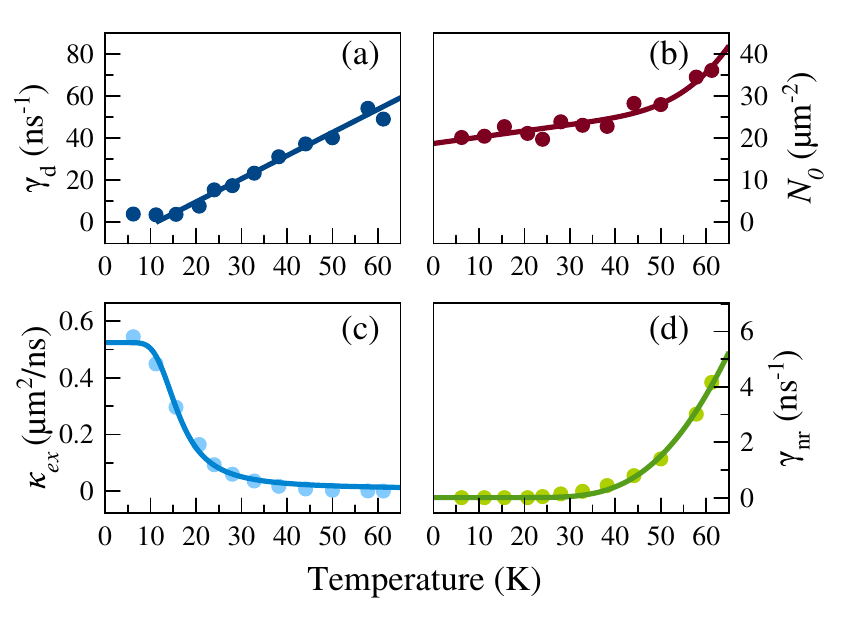}
    \caption{Temperature dependencies of the model parameters (dots):
(a) $\gamma_d$, (b) $N_0$, (c) $\kappa_{ex}$, (d) $\gamma_{nr}$. 
Solid curves are the fits (see details in the text).} 
\label{fig:temp_params}
\end{figure}

The temperature dependencies of the model parameters obtained in the simulation are shown in Fig.~\ref{fig:temp_params}. The exciton dissociation rate $\gamma_d$ remains almost constant at low temperatures, $\gamma_d = 3.8$~ns$^{-1}$ for $T < 15$~K, and then it linearly increases with the slope constant $d\gamma_d/dT = 1.1$~ns$^{-1}/K$ (see panel (a) in Fig.~\ref{fig:temp_params}). 

The exciton density $N_0$ created by a laser pulse in the reservoir also increases with temperature, see panel (b) in Fig.~\ref{fig:temp_params}. We assume that this effect is caused by the increase of the absorption coefficient at the Xhh resonance. Indeed, the resonance is narrow at low temperature ($\hbar(\Gamma_0 + \Gamma) < 0.1$~meV) relative to the spectral half width at half maximum of the laser pulse (HWHM $\approx 1$~meV). The temperature-induced broadening of the resonance increases its overlap with the laser pulse thus increasing the absorption. The main reason for the broadening is probably the exciton-phonon scattering. As we already discussed above (see previous subsection), the broadening is well described by Eq.~(\ref{Eq-phonons}). We found that this formula well fits dependence $N_0(T)$ with the same $E_{LO}=36$~meV, see solid curve in Fig.~\ref{fig:temp_params}(b). 

The temperature dependence of the bimolecular constant $\kappa_{ex}$ describing  the coupling of free electrons and holes into excitons is shown in Fig.~\ref{fig:temp_params}(c). It can be fitted by the Arrhenius function~(\ref{Eq-Arrhenius}) with activation energy, $E_{\text{act}}=7$~meV, coinciding with the exciton binding energy (exciton Rydberg) in the QW under study. This is natural consequence of the temperature-induced shift of the dynamic equilibrium in the reservoir towards free carriers. Indeed, small $\kappa_{ex}$ at temperatures $T> 30$~K means that the probability of the free carriers to couple into excitons is very small. Negligibly small areal density of excitons at delay $ \tau > 0.1$~ns observed in Fig.~\ref{fig:temp}(b) is explained by this fact. 

The rate of the nonradiative carrier relaxation as a function of temperature is well modeled by function,
$$
\gamma_{nr}(T)=\gamma_{nr}(0) \exp(-E_{\text{act}}/kT),
$$
see Fig.~\ref{fig:temp_params}(d). The activation energy for this process is relatively large, $E_{\text{act}}=23$~meV. It is comparable to the sum of localization energies, $\Delta E_e + \Delta E_{hh} \approx 30$~meV, of the electron and the hole in the QW under study~\cite{Kurdyubov-PRB2021}. Hence we can assume that the main mechanism of the nonradiative losses in the reservoir is related to the temperature-induced ejection of free carriers into the barrier layers followed by their motion across the heterostructure and radiative or nonradiative recombination in other layers. Similar mechanism of nonradiative losses of free carriers is extensively discussed in Ref.~\cite{Bacher-PRB1993} for shallow InGaAs/GaAs QWs. 

Equations~(\ref{Eqn-system1T}) contain one more free parameter, $\kappa_c$, describing the scattering of the reservoir excitons into light cone. We have found  that this parameter weakly depends on temperature and, therefore, we have fixed its value, $\kappa_c = 0.025$~$\mu$m$^2/$ns. This value agrees with that obtained in the pump power dependencies of the exciton dynamics discussed the next section.

\begin{figure}
\includegraphics[width=1\columnwidth]{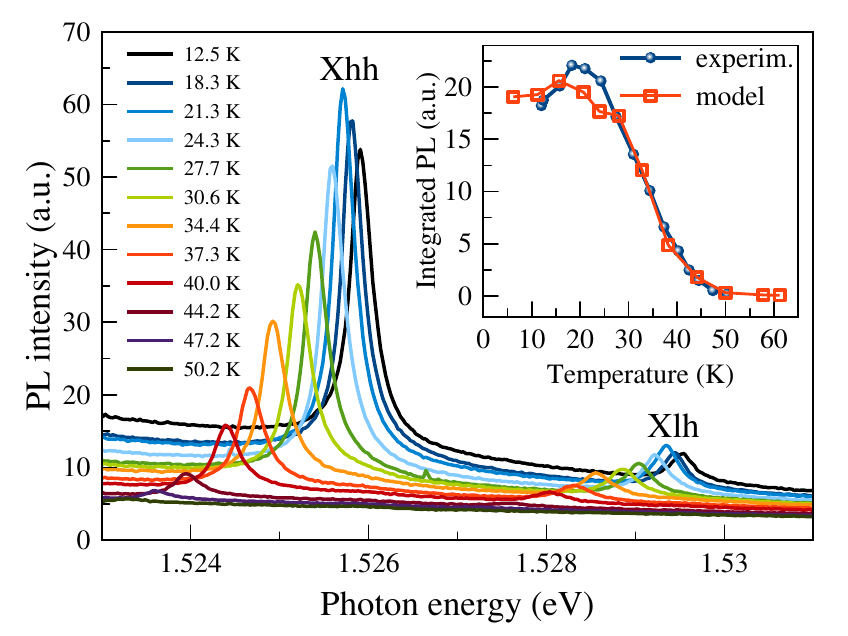}
\caption{PL spectra at different temperatures. Inset: temperature dependence of integral PL intensity of the Xhh excitons: experiment (blue circles) and model (squares). }
\label{PL-T}
\end{figure}

To verify the model described by Eqs.~(\ref{Eqn-system1T}), we have studied the PL yield of the Xhh exciton as a function of temperature. The PL was excited by a CW Ti:sapphire laser whose photon energy was tuned slightly above the Xlh resonance (to the ``step2'' spectral region~\cite{Kurdyubov-PRB2021}). Such excitation create free electrons and holes in the QW layer. Figure~\ref{PL-T} shows examples of PL spectra measured at different temperatures  and the obtained temperature dependence of the PL yield that is the integral PL intensity at the Xhh exciton transition. 
 
As it is already discussed in Ref.~\cite{Kurdyubov-PRB2021}, the PL can be modeled by the last term of the second equation in the system~(\ref{Eqn-system1T}), 
\begin{equation}
I_{PL}(T) = I_0 \kappa_c \int\limits_0^{\infty} n(t, T) N_x(t, T)\, dt,
\label{PL}
\end{equation}
where $I_0$ is a scaling factor.

The dependence $I_{PL}(T)$ calculated using densities $n(t, T)$ and $N_x(t,T)$ obtained for each temperature is shown in Fig.~\ref{PL-T}. As seen, the calculated dependence well reproduces the experimentally measured dependence $I_{PL}(T)$ with no fitting parameters except a scaling factor. So, the temperature-induced decrease of the PL yield is also explained by the ejection of free carriers from the QW into the barrier layers. 

\section{Power dependencies}
\label{sec:power_dep}

\begin{figure}
   \includegraphics[width=1\columnwidth]{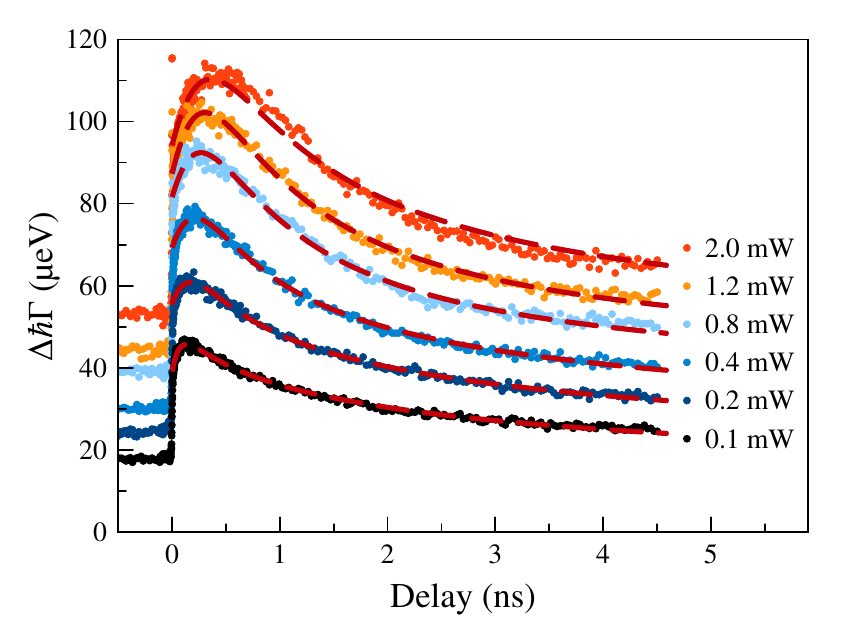}
   \caption{Dynamics of the nonradiative broadening of the Xhh resonance under excitation into Xlh at different pump powers in the range from 0.1 to 2 mW. Experimental data are shown by dots. Dashed curves show the results of modeling by equations~(\ref{Eqn-system1}). $T = 4.5$~K.}
   \label{Dynamics_power_Xlh}
\end{figure}

We have studied the dynamics of nonradiative broadening of the Xhh resonance at the excitation to the Xlh resonance with different powers at T = 4.5~K. Examples of the experimental data are shown in Fig.~\ref{Dynamics_power_Xlh}. As seen from the figure, the magnitude of the nonradiative broadening sublinearly increases with power. The broadening increases only 2.5 times when the power increases from 0.1~mW to 2.0~mW, that is 20 times. 

The dynamics of nonradiative broadening has been modeled using Eqs.~(\ref{Eqn-system1}). The characteristic rates of the processes considered in the model are used as fitting parameters. The model well describes the dynamics for all the powers used in the experiment, see dashed curves in Fig.~\ref{Dynamics_power_Xlh}. This fact allows one to reliably determine values of the parameters and to analyze their dependence on the excitation power. 

\begin{figure}
   \includegraphics[width=1\columnwidth]{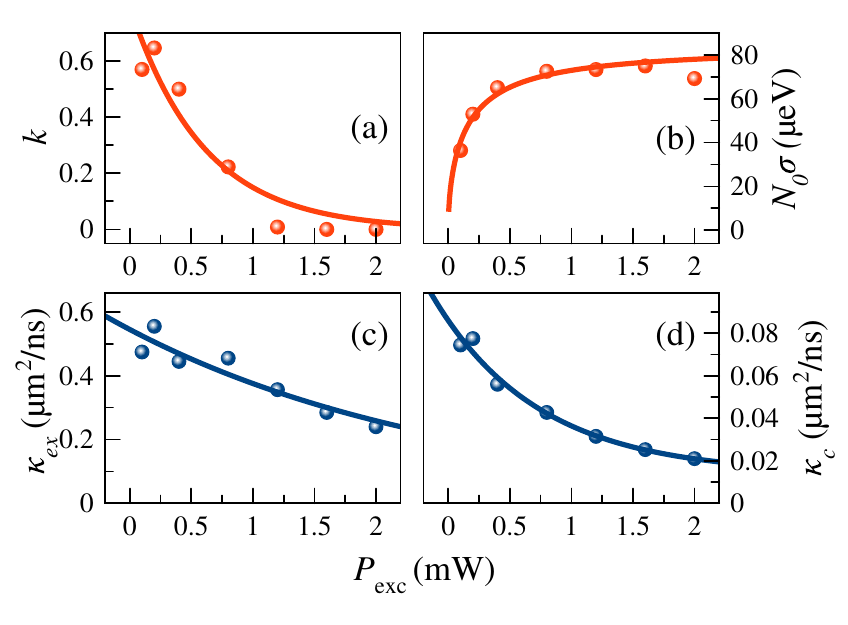}
   \caption{Power dependencies of fitting parameters of the system~(\ref{Eqn-system1}) (points). (a) the fraction $k$ of ``cold'' excitons scattered into the reservoir just after the arrival of the pump pulse; (b) the product of the areal density $N_0$ of the Xlh excitons scattered to the reservoir after each pulse and the cross-section $\sigma$ of the exciton-exciton scattering;  (c) the bimolecular rate constant $ \kappa_{ex} $ of the electron-hole coupling into the reservoir excitons; (d) the bimolecular rate constant $ \kappa_{c} $ of the exciton scattering into the light cone. The solid fitting curves are described in the text.}
   \label{Dynamics_params_Xlh}
\end{figure}

The first parameter is the fraction of ``cold'' excitons ($k$), which are scattered into the reservoir just after the excitation pulse. As Fig.~\ref{Dynamics_params_Xlh}(a) shows, fraction $k$ rapidly decreases almost down to zero at $P_{\text{exc}} > 1$~mW. Its power dependence can be approximated by a phenomenological function, $k(P_{\text{exc}}) = k_0\exp(-P_{\text{exc}}/P_{0})$, with $k_0 = 0.8$ and $P_0 = 0.6$~mW. It is intuitively clear that strong excitation should warm the reservoir because it creates Xlh excitons with excess energy of about 3~meV relative to the Xhh excitons (see inset in Fig.~\ref{fig:dynamics_T670}).  Correspondingly, the fraction $k$ of ``cold'' excitons should decrease with increasing power and the fraction $(1-k)$ of ``hot'' excitons should increase accordingly. Recall that the ``hot'' excitons in our model dissociate into free electrons and holes while the ``cold'' excitons are accumulated in the reservoir. The dissociation time of ``hot'' excitons is found to be $\tau_d = 1/\gamma_d \approx 400$~ps at the pump power $P_{\text{exc}} > 0.5$~mW. At lower pump powers, the accuracy of determination of $\tau_d$ from the experimental data is insufficient to obtain a reliable value due to small magnitude of the photoinduced nonradiative broadening.


Panels (c) and (d) in Fig.~\ref{Dynamics_params_Xlh} show the power dependencies of the rates of bimolecular processes $ \kappa_{ex} $ and $ \kappa_c $, respectively. Solid curves are fits by function $ y = c \exp (-P_{exc} / P_0) + d $, with parameters $ c = 0.55 $~$ \mu $m$^2 / $ns, $ P_0 = 2.7 $~mW, $ d = 0 $ for $ \kappa_ {ex} $; $ c = 0.072 $~$ \mu $m$^2 / $ns, $ P_0 = 0.86 $~mW, $ d = 0.014 $~$ \mu $m$^2 / $ns for $ \kappa_{c} $. Behavior of these parameters also depends on the effective reservoir temperature as it will be discussed below.

\begin{figure}
   \includegraphics[width=1\columnwidth]{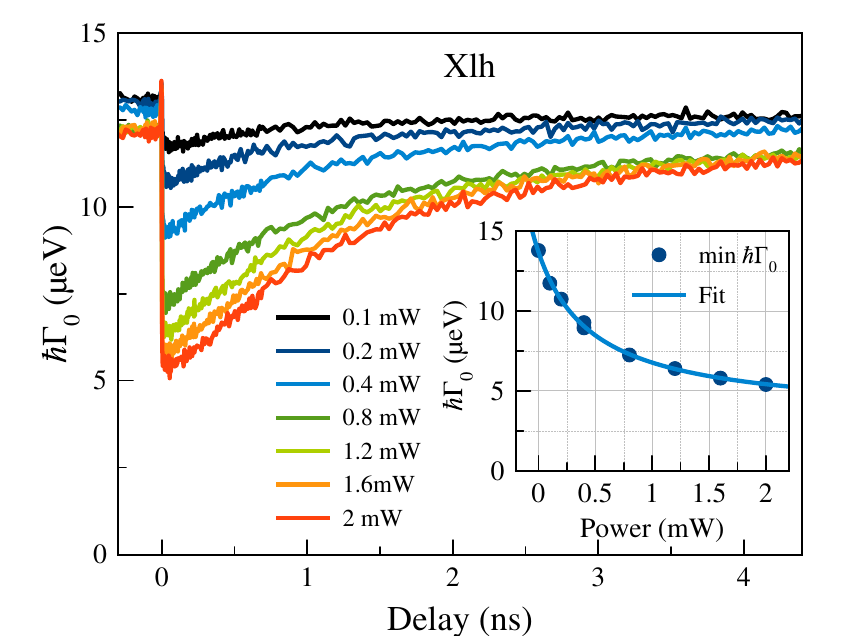}
   \caption{Dynamics of radiative broadening of the Xlh resonance obtained under excitation into this resonance at different excitation powers. Inset shows the power dependence of minimum value of the broadening (points) and its fit by function $f(P_{\text{exc}})= A/(1+a P_{\text{exc}})+d$ with parameters: $A = 10.3$~$\mu$eV, $a = 2.06$~mW$^{-1}$, $d = 3.4$~$\mu$eV.}
   \label{Dynamics-Xlh-Gamma0}
\end{figure}

An important parameter of the model is $N_0$, that is the number of excitons (per unit area) scattered after excitation by each pulse. Its power dependence is shown in panel (b) of Fig.~\ref{Dynamics_params_Xlh}. Strictly speaking, this parameter cannot be determined independently of the parameter $ \sigma$, since only their product determines the broadening of the exciton resonance measured in the experiment, see Eq.~(\ref{scattering}). The scattering cross-section $ \sigma$ vary with the power~\cite{Gribakin-PRB2021}. The power dependence of $ N_0 \sigma $ is shown in Fig.~\ref{Dynamics_params_Xlh}(b). As seen it is strongly sublinear that reflects the sublinearity of magnitude of the dynamic curves shown in Fig.~\ref{Dynamics_power_Xlh}.  We approximated the dependence of product $ N_0 \sigma $ on $ P_{\text{exc}} $ by a phenomenological function $ f(P_{\text{exc}}) =A P_{\text{exc}}/(1+a P_{\text{exc}})$ with parameters $ A = 720$~$ \mu$eV and $ a =8.9$~mW$^{-1}$. This phenomenological function reflects the power dependence of respective physical quantities: 
\begin{equation}
N_0 \sigma  = P_{\text{exc}} \alpha(P_{\text{exc}}) \sigma(P_{\text{exc}}). 
\end{equation}
The absorption coefficient $\alpha$ for the Xlh transition decreases with power. In the case of the large nonradiative broadening of exciton resonance, the absorption coefficient is proportional to the {\emph{radiative}} broadening of the resonance $\hbar \Gamma_0$, describing the exciton-light coupling strength~\cite{Ivchenko-book}. The dynamics of $\hbar \Gamma_0$ for the Xlh resonance obtained at several excitation powers is shown in Fig.~\ref{Dynamics-Xlh-Gamma0}. The data are measured in the same experimental series as shown in Fig.~\ref{Dynamics_power_Xlh} for the nonradiative broadening of the Xhh resonance. 

As seen in Fig.~\ref{Dynamics-Xlh-Gamma0}, the radiative broadening and, correspondingly, the absorption coefficient drop by more than 2 times when the Xlh resonance is excited with strong enough power $P > 1$~mW. It is noteworthy that the absorption coefficient is then being restored for a long time, within a few nanoseconds. This is so-called ``state filling effect'', which is extensively studied for the quantum dot structures~\cite{Raymond-PRB1996}. The transfer of a significant number of electrons from the valence band to the conduction band leads to bleaching of the optical transition since fewer electrons remain in the ground state.  The depopulation of the initial (``vacuum'') state of the structure and transfer of electron-hole pairs into the long-lived nonradiative reservoir explains the long lifetime of the effect. 

The dependence of the scattering cross-section $ \sigma$ on the power is caused by the dependence of the exciton--exciton and exciton--carrier interaction on the momentum $q$ of interacting quasiparticles. The microscopic modeling~\cite{Gribakin-PRB2021} shows that the energy of the interaction decreases with $q$ and becomes close to zero at $q a_B \approx 1.5$ where $a_B$ is the exciton Bohr radius.  In our case, bright excitons with zero momentum interact with dark excitons and free carriers in the reservoir with finite momenta $q$. The increase of the areal density of quasiparticles in the reservoir effectively increases their average momentum and, correspondingly, decreases their interaction. This effect can be described introducing the density dependent scattering cross-section $\sigma (n)$~\cite{Gribakin-PRB2021},
\begin{equation}
\sigma(n) = \frac{\sigma_0}{1 + \sqrt{n/n_0}}.
\label{sigma-n}
\end{equation}
Here $\sigma_0$ is the scattering cross-section at the infinitely small density and $n_0$ is the density when the cross-section drops twice. The modeling of Ref.~\cite{Gribakin-PRB2021} shows that $n_0 \approx 7$~$\mu$m$^{-2}$ for the X-X and X-e scattering and even smaller for the X-h scattering. These values are small compared to densities achieved in our experiments, see Fig.~\ref{Dynamics_params_Xlh}(b), taking into account that $\sigma \le 3$~$\mu$m$^2\cdot\mu$eV. Correspondingly, the scattering cross-section $\sigma$ can considerably decrease with the excitation power rise. 

\begin{figure}
   \includegraphics[width=1\columnwidth]{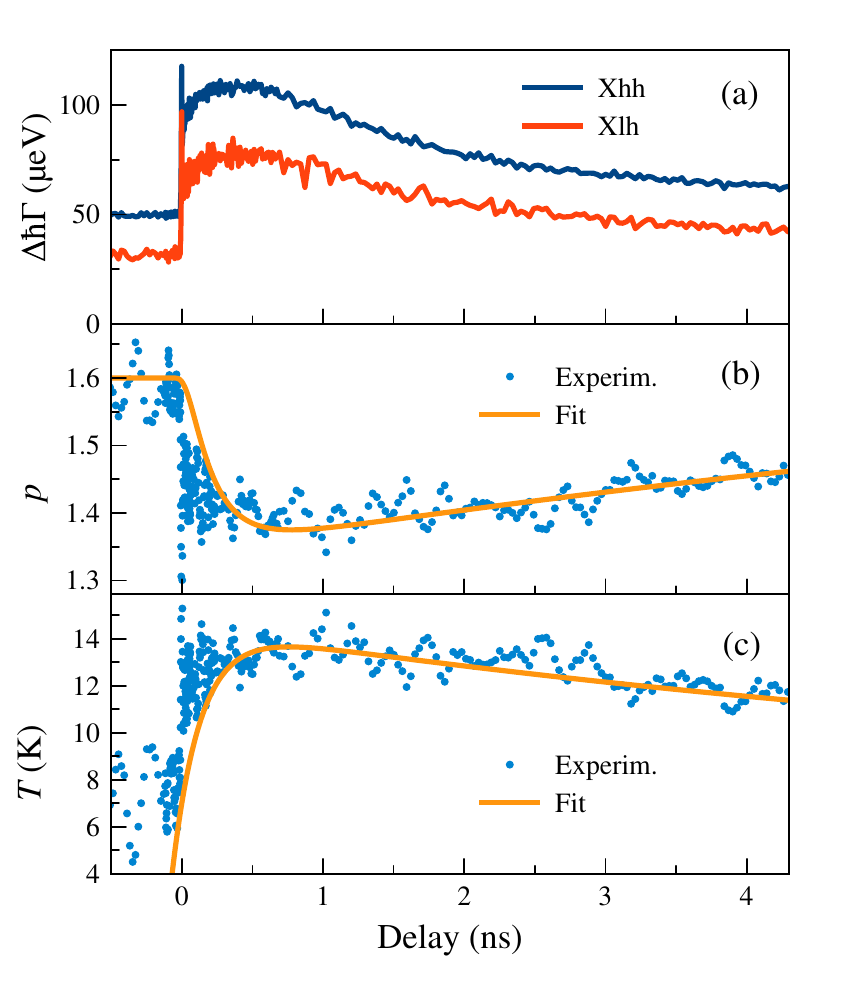}
   \caption{(a) Dynamics of the photoinduced part of nonradiative broadening of the Xhh and Xlh excitons measured at the excitation power $P = 1.6$~mW. (b) Ratio of the curves shown in panel (a) (noisy curve). Smooth curve is the fit (see the text for details). (c) Dynamics of reservoir temperature extracted from the experiment (noisy curve). Smooth curve is the fit by function~(\ref{Tt}) with parameters: $T_0 = 5.5$~K, $\Delta T = 7$~K, $t_1 = 7$~ns, and $t_2 = 0.15$~ns. Sample temperature $T = 4.5$~K.}
   \label{Dynamics_ratio}
\end{figure}

The average momentum $q$ can also increase because of warming of the reservoir, either due to strong excitation (see Fig.~\ref{Dynamics_ratio}) or due to heating the sample. Estimates show that  $q a_B \approx 1.5$ and, correspondingly, the interaction strength becomes close to zero at the reservoir temperature $T \approx 50$~K.
 
The increase of the reservoir effective temperature at sufficiently strong pumping can be evaluated using a ``thermometer'' based on the ratio of dynamic curves of the nonradiative broadening of the Xhh and Xlh excitons. An example of such curves measured at $P = 1.6$~mW is shown in Fig.~\ref{Dynamics_ratio}(a). If the reservoir temperature were very low the Xhh exciton broadening would be twice as large as the Xlh exciton broadening as it was already discussed in the previous section. Figure \ref{Dynamics_ratio}(b) shows such ratio. Although this ratio is relatively noisy, it is significantly lower than two. Moreover, it further decreases when the sample is excited by a strong pump pulse. This points out the increase in the reservoir temperature. 

We model the dynamic of the ratio, $p(t)$, by the function~(\ref{pT}). The time dependence of the effective temperature is simulated by a two-exponential function, 
\begin{equation}
T(t) = T_0+\Delta T \left[\exp(-t/t_1) - \exp(-t/t_2)\right].
\label{Tt}
\end{equation}
This dependence describes establishing the effective reservoir temperature with characteristic time $t_2$ and its relaxation with characteristic time $t_1$. Equations~(\ref{pT}, \ref{Tt}) are used to simulate temporal dependence of the ratio shown in Fig.~\ref{Dynamics_ratio}. In the simulation, parameters of the Arrhenius function are taken from the caption of Fig.~\ref{T670_temp}(f), $p_0 = 0.61$. Parameters of function~(\ref{Tt}) are: $t_1 = 7$~ns, $t_2 = 0.15$~ns, $T_0 = 5$~K, $\Delta T = 8$~K. So, the dynamic equilibrium in the reservoir is achieved within time $t_2 = 0.15$~ns. The reservoir temperature increases up to $T_0 + \Delta T = 13$~K and then slowly relaxes down to $T_0 = 5$~K with decay time $t_1 = 7$~ns.

To make the temperature change more visual, we transformed the dependence $p(t)$ obtained from the experiment using the Arrhenius function~(\ref{Eq-Arrhenius}):
\begin{equation}
T(t) = -\frac{E_{\text{act}}/k_B}{\ln\!\left[\frac{1}{a}\left(\frac{A_0}{p(t)-p_0}-1\right)\right]}.
\label{Tt-exp}
\end{equation}
The result of such transformation of the experimental data is shown in panel (c) of Fig.~\ref{Dynamics_ratio}. It clearly shows the reservoir temperature rise from $T_0 \approx 5.5$~K before the excitation pulse arrival (the negative delay) up to approximately 12~K after the pulse. The time dependence of the effective reservoir temperature can be fitted by function~(\ref{Tt}) with parameters given in the caption of Fig.~\ref{Dynamics_ratio}. Some difference of these parameters from those for the curve in panel (b) is caused by the nonlinear transformation~(\ref{Tt-exp}) and the specifics of the least-square fitting procedure.

An increase in temperature shifts the thermodynamic equilibrium excitons~$ \leftrightarrow $~free charge carriers towards the latter. This process is widely discussed in many papers (see~\cite{Colocci-JAP1990, Szczytko-PRL2004, Deveaud-ChemPhys2005, Kaindl-PRB2009}), where it is reported that $ \kappa_{ex} $ decreases with increasing temperature. The temperature rise also causes an increase in the average wave vector of excitons in the reservoir. In other words, they move away from the light cone in the reciprocal space, and their scattering into radiative states becomes less probable. This can explain the decrease of $ \kappa_c $ with increasing pump power shown in Fig.~\ref{Dynamics_params_Xlh}(d).


\section{Concluding remarks}

The study performed shows that the exciton system in quantum wells behaves as a highly nonlinear medium. The nonlinear optical response of bright excitons (Fig.~\ref{Dynamics_power_Xlh} and \ref{Dynamics-Xlh-Gamma0}) is caused by their interaction with dark excitons and free charge carriers, which form the nonradiative reservoir. The areal density and the dynamics of quasi-particles in the reservoir are controlled by several processes. The reservoir is populated via scattering of the photocreated excitons into the states with large in-plane wave vector exceeding the wave vector of light. The efficiency of the scattering nonlinearly depends on the excitation power due to the bleaching of the exciton transition (Fig.~\ref{Dynamics-Xlh-Gamma0}). The efficiency also depends on the areal density of the quasi-particles interacting with bright excitons. The reservoir excitons do not interact with light and live for tens of nanoseconds in the structure under study. 

The exciton density in the reservoir depends on the depleting processes. In high-quality heterostructures the main mechanism of depletion of the reservoir at low temperatures is the scattering of excitons into the light cone, where they quickly recombine with the emission of photons. The exciton--carrier scattering dominates over other depleting processes. 

Increasing of the reservoir density gives rise to a decrease of the scattering cross-section $\sigma$. This phenomenon is associated with the increase of the mean wave-vector $q$ of quasi-particles in the reservoir and corresponding decrease in the energy of their exchange interaction~\cite{Gribakin-PRB2021}. 

Strong optical excitation leads to heating of the exciton reservoir and activation of other processes. To visualize the heating effect, we investigated the temperature dependence of the exciton dynamics in the reservoir. The heating is accompanied by a shift in the dynamic equilibrium of excitons $\leftrightarrow$ free charge carriers towards the latter. This leads to an acceleration of the scattering of excitons from the light cone and a reduction in their lifetime in the reservoir. In addition, the states of the light-hole excitons, which are separated from the states of the heavy-hole excitons, are populated. As a result, the interaction of excitons in the reservoir with the light-hole and heavy-hole excitons in the light cone levels out. In the experiment, this is observed as equalization of the photoinduced nonradiative broadening of the light-hole and heavy-hole exciton resonances (Fig.~\ref{T670_temp}). We use this effect to measure the effective temperature of the exciton reservoir at strong pumping. The ratio of the broadenings can serve as a
kind of thermometer for the reservoir. We found that, at the excitation power density of about 10~W/cm$^{2}$, the reservoir temperature increases by about 7~K within 0.15~ns and then slowly relaxes down with characteristic time 9~ns (Fig.~\ref{Dynamics_ratio}).

At high enough reservoir temperatures ($>30$~K), when the dynamic equilibrium of excitons $\leftrightarrow$ free charge carriers is shifted towards the latter, a channel of the nonradiative losses is activated. This is the thermally-induced escape of free charge carriers from the QW into the barrier layers. The losses give rise to the experimentally observed decrease of the PL yield (Fig.~\ref{PL-T}).

We developed a model of dynamic processes in the exciton system. The model considers several processes: the excitation-induced population of the reservoir by the nonradiative excitons and free charge carriers, the coupling of free electrons and holes into excitons, the depletion of the reservoir via exciton scattering into the light cone induced by the exciton--free carrier interaction, the thermally activated nonradiative losses of the charge carriers (see Fig.~\ref{fig:scheme1}). To simplify the model, we did not take into account the pump-induced change in the effective temperature of the reservoir. The model allowed us to quantitatively describe the dynamics of the exciton and carrier densities in the reservoir (Fig.~\ref{fig:temp}) and to determine the temperature and power dependencies of parameters of the model (Figs.~\ref{fig:temp_params} and \ref{Dynamics_params_Xlh}). The model also quantitatively describes the decrease of the PL yield with no fitting parameters.


\begin{acknowledgments}
The authors acknowledge Saint-Petersburg State University for the financial support in the framework of grant No. 91182694. The authors thank Recourse Center “Nanophotonics” SPbU for the heterostructure studied in this work. I.V.I. acknowledges the Russian Science Foundation (Grant No. 19-72-20039) for the financial support of the theoretical analysis performed in this work. 
\end{acknowledgments}

\end{document}